\documentclass{article}
\usepackage{amsmath,amssymb}
\usepackage[dvips]{graphicx}

\newtheorem{theorem}{Theorem}
\newtheorem{lemma}{Lemma}

\newtheorem{corollary}{Corollary}

\newtheorem{remark}{Remark}

\newcommand{\bs}[1]{\boldsymbol{#1}}

\title{Localization of quantum walks \\ induced by recurrence properties of random walks}
\author{
Etsuo Segawa\footnote{E-mail: segawa@stat.t.u-tokyo.ac.jp} \\
{\small {\it Department of Mathematical Informatics,} }\\ 
{\small {\it University of Tokyo, Bunkyo, Tokyo, 113-8656, Japan.} }
}
\date{}
\begin{document}
\maketitle
\noindent \textbf{Abstract}: 
We study a quantum walk (QW) whose time evolution is induced by a random walk (RW) first introduced by Szegedy (2004). 
We focus on a relation between recurrent properties of the RW and localization of the corresponding QW. 
We find the following two fundamental derivations of localization of the QW. 
The first one is the set of all the $\ell^2$ summable eigenvectors of the corresponding RW. 
The second one is the orthogonal complement, whose eigenvalues are $\pm 1$, of the subspace induced by the RW. 
In particular, as a consequence, for an infinite half line, we show that 
localization of the QW can be ensured by the positive recurrence of the corresponding RWs, 
and also that the existence of only one self loop affects localization properties. 
\section{Introduction}
A frame work for quantum search algorithms by using the coin-shift QW on a spatial structure 
was constructed by \cite{SKW,AKR,Ambainis} for example. 
Its one step time evolution $U$ is given by $U=SC$: 
the first operator $C$ is so called coin flip assigning a unitary operator called quantum coin (in many case, the Grover operator) to each vertex, 
where the dimension of the quantum coin is the degree of the corresponding vertex, 
and the second operator $S$ is called shift which moves a quantum particle to its neighbor. 
The paper \cite{Ambainis_Rev} is one of the nice reviews on these works. 
Szegedy \cite{Sze} proposed a generalized QW whose time evolution comes from a corresponding RW on a bipartite graph. 
This walk is called bipartite walk or the Szegedy walk. 
Let the time evolution of the Szegedy walk and a coin-shift QW be $W$, and $U$, respectively. Then as mentioned by \cite{MNRS1} for example, 
an appropriate choice of the coin flip operator of $U$ gives $W=U^2$. The eigensystem of $W$ is explicitly obtained in \cite{Sze}. 
In \cite{MNRS1,MNRS2}, they blush up the quantum search algorithm and a notion of quantum hitting time of \cite{Sze} with the walk. 

In these works, the eigensystem of the subspace induced by the spectrum of the corresponding RW, named $\mathcal{H}^{(R)}$ in this paper, 
plays an important role: the spectral gap of the corresponding RW determines the quality of the quantum search algorithm. 
Moreover graphs treated in these studies are finite because of its algorithmic application. 
Since $\mathcal{H}^{(R)}$ is invariant under the action of $U$ and $W$, an appropriate initial state gives us considerations 
outside of its orthogonal complement space ${\mathcal{H}^{(R)}}^{\bot}\equiv \mathcal{H}^{(S)}$. 
In this paper, we take an initial state which is a convex combination of $\bs{u}\in \mathcal{H}^{(R)}$ and $\bs{v}\in \mathcal{H}^{(S)}$ to 
get some effects of the subspace $\mathcal{H}^{(S)}$ on behaviors of the QW, 
and also treat a simple infinite graph, that is, a half line with self loops. 
We focus on the time averaged limit measure. 
Due to its infiniteness of the system, the distributivity of the time averaged limit measure is not ensured. 
In fact, except in the case of positive recurrent case on the half line, the time averaged limit measure of the RW becomes the null measure. 
In this paper, if we can take an appropriate initial state so that the time averaged limit measure of the QW is positive, 
then we say that localization of the QW occurs. We show that all the $\ell^2$ summable eigenvectors of the RW and 
the orthogonal projection onto $\mathcal{H}^{(S)}$ of the initial state are essential to the derivations of localization of the QW. 
Off course, the existence of a finite number of self loops has little influence to the behavior of the RW. 
However as we will see, since the number of the self loops determines the dimension of the subspace of $\mathcal{H}^{(S)}$ in the QW, 
the behavior, especially localization property, is quite sensitive to the self loops. 
Actually, adding only one self loop to even a transient RW gives a change from a QW without localization to a QW which exhibits localization. 

This paper is organized as follows. 
In Sect. 2, we give a review on the Szegedy walk and clarify the eigensystem of the time evolution $U$ on graph $G(V,E)$, 
where $V$ is the set of vertices, and $E$ is the set of edges. 
This paper treats graphs with self loops. We denote the set of of self loops as $\mathcal{S}\subset E$. 
Since the time evolution of \cite{Sze} $W$, is expressed by $W=U^2$, 
the statement of part 4 in Theorem 1 of \cite{Sze} implies that the possibilities of eigenvalues of $\mathcal{H}^{(S)}$ are $\{1\}$, $\{-1\}$, and $\{1,-1\}$. 
In this section, we find that the eigenvalues of $\mathcal{H}^{(S)}$ are $\pm 1$ with multiplicities $|E|-|V|+m(1)-|\mathcal{S}|$ and $|E|-|V|+m(-1)$, respectively. 
Here $m(\pm 1)$ are the multiplicities of the eigenvalues ${\pm 1}$ of the corresponding RW. 
Section 3 proposes a relation between recurrence property of the RW and localization property of the QW. 
It is obtained that the time averaged limit measure of the QW is bounded below by the stationary distribution of the positive recurrent RW. 
We also see that this localization comes from $\mathcal{H}^{(R)}$. 
On the other hand, we find that 
the effect of $\mathcal{H}^{(S)}$ to the localization property on an infinite half line 
is determined by the number of self loops and the recurrence property of the corresponding RW. 
More precisely, the support of localization in $\{0,1,2,\dots\}$ with a special initial state 
depends on the recurrence property of the corresponding RW with finite self loops. 
Finally, we give a discussion. 
\section{Review on the Szegedy walk}
\subsection{Definition of the quantum walk}
In this paper, we denote the Hilbert space of square-summable functions on a countable set $X$ by $\ell^2(X)$. 
The inner product is defined by
\[ \langle \bs{f}, \bs{g} \rangle=\sum_{x\in X}\overline{\bs{f}(x)}\bs{g}(x),\;\;\; \bs{f},\bs{g}\in \ell^2(X) \]
and the norm by $||\bs{f}||=\langle \bs{f},\bs{f} \rangle^{1/2}$. 
We define a complete orthonormal system of $\ell^2(X)$ by $\{\bs{\delta}_x\}_{x\in X}$ with 
\[ \bs{\delta}_x(y)=\begin{cases} 1 & \text{: $x=y$,}\\ 0 & \text{: $x\neq y$.} \end{cases} \]
The graph $G(V,E)$ treated here is connected and all the degrees of vertices are finite. 
In this paper, we treat graphs with self loops. The self loop means $(u,v)\in A(G')$ with $u=v$. 
We denote $\mathcal{S}$ as the set of all the self loops. 
At first, we consider an ergodic RW on the graph $G(V,E)$. 
Let $M$ be the stochastic matrix of the RW, $(M)_{u,v}=p_{u,v}$ with $\sum_{u\in E(G)}p_{u,v}=1$. 
(a walker located in position $v\in V(G)$ jumps to the neighbor $u$ with probability $p_{u,v}$. ) 
We also give a self adjoint matrix $J$ defined by $(J)_{u,v}=\sqrt{p_{u,v}p_{v,u}}$ which plays an important role in this paper. 
Recall that if $M$ has a reversible measure $\boldsymbol{\pi}(j)=\pi_j$, then $p_{ij}\pi_j=p_{ji}\pi_i$. 
This relation is called ``detailed balance condition". 
When $\pi_j \neq 0$ for every $j$, then $J=D^{-1/2}MD^{1/2}$, 
where $D$ is a diagonal matrix with $(D)_{jj}=\pi_j$. 
In the following, we should review recurrence properties of RW, which appear again to the localization of QW. 
For more details, see \cite{Feller,Schinazi} for example. 
If the probability that a number of revisits to the starting place is infinite equals to $1$, then we say that the walk is recurrent, otherwise transient. 
Moreover if the mean first hitting time in the recurrent walk is finite, then we call it positive recurrent, otherwise null recurrent. 
\noindent \\
\noindent The symmetric oriented graph of $G(V,E)$ is a digraph $G'(V,A)$ whose vertices and arcs are defined as follows: \\
$V(G')=V(G)$ and 
$\{u,v\}\in E(G)$ iff $(u,v), (v,u)\in A(G')$, 
where $(u,v)$ is the arc from the vertices $v\in V(G')$ to $u\in V(G')$. 
From now on, we give a QW corresponding to the above RW. 
The total space of the QW is $\mathcal{H}\equiv \ell^2(A)$ whose orthonormal basis are associated with the arcs of its symmetric oriented graph $G'$, 
that is, 
\[ \mathcal{H}\equiv \ell^2(A)=\mathrm{span}\{ \bs{\delta}_{(u,v)}: (u,v)\in A(G')\}. \] 
Note that the dimension of our quantum walk is $|A|=2|E|-|\mathcal{S}|$. 
The time evolution $U=SC$ is defined as follows:
\begin{itemize}
\item Shift $S$: 
For any $\boldsymbol{\delta}_{(u,v)}$ with $(u,v)\in A(G')$, 
\begin{equation}
S\boldsymbol{\delta}_{(u,v)}=\boldsymbol{\delta}_{(v,u)}. 
\end{equation}
This is a permutation on $A(G')$ called ``flip flop" \cite{AKR}. 
\item Coin flip $C$: 
Let $\boldsymbol{a}_u$ and $\boldsymbol{b}_u$ $(u\in V(G'))$ be 
\[ \boldsymbol{a}_u=\sum_{v: (v,u)\in A(G')}\sqrt{p_{v,u}}\boldsymbol{\delta}_{(v,u)},\;\;\boldsymbol{b}_u=S\boldsymbol{a}_u. \]
We call $\boldsymbol{a}_u$ and $\boldsymbol{b}_u$ ``incidence vector of $u$" and ``swapped incidence vector of $u$", respectively in this paper. 
Denote $\Pi_A$ and $\Pi_B$ as the orthogonal projections onto the subspaces of $\mathcal{H}_\mathcal{A}=\mathrm{span}\{\boldsymbol{a}_u: u\in V(G')\}$ and 
$\mathcal{H}_\mathcal{B}=\mathrm{span}\{\boldsymbol{b}_u: u\in V(G)\}$, respectively. Then the coin flip $C$ is defined by 
\begin{equation} C=2\Pi_A-I. \end{equation}
\end{itemize}
Noting $S^2=I$, the one-step Szegedy walk, $W$, is equivalent to $W=U^2=\mathrm{ref}_B\mathrm{ref}_A$, where $\mathrm{ref}_A=2\Pi_A-I(=C)$ and 
$\mathrm{ref}_B=2\Pi_B-I$. 
An equivalent expression for the QW is that: 
a quantum particle on $v\in V$ at the present time step which was in $v\in V$ at the previous time step, moves to its neighbor, $w\in V$, with amplitude 
$2\sqrt{p_{w,v}p_{u,v}}-\delta_{w,u}$ at the next time step. 
So only if a quantum particle returns back to the same place, an extra term $-1$ appears in the amplitude associated with its moving. 
When the corresponding RW is symmetric in that $p_{u,v}=1/\mathrm{deg}(v)$ for every $v\in V$, where $\mathrm{deg}(v)$ is the degree of $v$, 
the QW becomes so called the Grover walk which has been intensively investigated. \\
For $\bs{\Psi}_0\in \ell^2(A)$ with $||\bs{\Psi}_0||^2=1$, let $X_t^{(\Psi_0)}$ be a random variable which is determined as follows. 
\[ P(X_t^{(\Psi_0)}=u)=\sum_{v:(v,u)\in A(G)}\left|\langle \bs{\delta}_{(v,u)},U^t\boldsymbol{\Psi}_0\rangle \right|^2. \]
The random variable $X_t^{(\Psi_0)}$ corresponds to a position when the QW with the initial state ${\Psi_0}$ is measured at time $t$. 
\subsection{Eigensystem of the QW}
Let $A$ be a weighted incidence matrix of $G$ defined by 
\[ A=\sum_{u\in V(G)} \boldsymbol{a}_u \boldsymbol{\delta}_u^{\dagger}. \]
Also define $B$ as 
\[ B=\sum_{u\in V(G)} \boldsymbol{b}_u \boldsymbol{\delta}_u^{\dagger}. \]
\begin{remark}\label{shiftoperator}
The eigenvalues of the shift operator $S$ are $\pm 1$ with multiplicities $|E|$, and $|E|-|\mathcal{S}|$, respectively. 
The eignespaces of eigenvalues $1$ and $-1$ for $S$ are described by 
\begin{align}
\mathcal{M}_+ &= \mathrm{span}\left\{\{ \bs{\delta}_{(u,v)}+\bs{\delta}_{(v,u)}: (u,v)\in A(G') \mathrm{\;with\;} u\neq v \}\cup \{ \bs{\delta}_{(u,v)}: (u,v)\in \mathcal{S} \}\right\}, \\
\mathcal{M}_- &= \mathrm{span}\{ \bs{\delta}_{(u,v)}-\bs{\delta}_{(v,u)}: (u,v)\in A(G') \mathrm{\;with\;} u\neq v \},
\end{align}
respectively. 
\end{remark}
The ``correspondence" between the RW and the QW derives from the following lemma. 
\begin{lemma}\label{acco1}\noindent \\
\begin{enumerate}
\item For $x\in [-1,1]$, let $\boldsymbol{\mathfrak{p}}(x)\in \ell^2(V)$ satisfy the following eigenequation: 
\begin{equation}\label{yoshio}
J\boldsymbol{\mathfrak{p}}(x)=x\boldsymbol{\mathfrak{p}}(x).
\end{equation}
Then we have 
\begin{equation}
U\boldsymbol{\mathfrak{q}}_{\pm}(x)=e^{\pm i\theta(x)}\boldsymbol{\mathfrak{q}}_{\pm}(x),
\end{equation}
where $x=\cos\theta(x)$, and $\boldsymbol{\mathfrak{q}}_{\pm}(x)\in \ell^2(A)$ with 
\begin{equation}\label{pero1}
\boldsymbol{\mathfrak{q}}_{\pm}(x)
	=\begin{cases}
        (I-e^{\pm i\theta(x)}S)A\bs{\mathfrak{p}}(x) & \text{: $x\notin \{\pm 1\}$, } \\
        A\bs{\mathfrak{p}}(x) & \text{: $x\in \{\pm 1\}$.}
         \end{cases} 
\end{equation}
\item If the orthogonal complement of $\mathcal{H}^{(R)}\equiv \mathrm{span}\{\mathcal{H}_A,\mathcal{H}_B\}$ exists, then 
$\mathcal{H}^{(S)}\equiv (\mathcal{H}^{(R)})^\bot$ is orthogonally decomposed into invariant subspaces with respect to the action of $U$, such that 
$\mathcal{H}^{(S)}=\mathcal{H}^{(S)}_+\oplus\mathcal{H}^{(S)}_-$, where 
for all $\bs{s}_\pm \in \mathcal{H}^{(S)}_\pm$, 
\[ U\bs{s}_+=\bs{s}_+,\;\;U\bs{s}_-=-\bs{s}_- \]
with multiplicities $|E|-|V|+m(-1)-|\mathcal{S}|$ and $|E|-|V|+m(1)$, respectively.
\end{enumerate}
\end{lemma}

\begin{remark}\label{katsumori1} 
If $|V|<\infty$, then $\mathrm{span}\{\bs{a}_u: u\in V(G)\}=\mathrm{span}\{A\bs{\frak{p}}(\lambda): \lambda\in \mathrm{spec}(J)\}$
since $\langle \bs{a}_u,\bs{a}_v \rangle=\delta_{u,v}$, $\langle \bs{b}_u,\bs{b}_v \rangle=\delta_{u,v}$ and from its finiteness of the system size, 
the eigenvectors of $J$ give the complete system of $\ell^2(V)$. 
Therefore we have 
\[\mathcal{H}^{(R)}=\mathrm{span}\{\bs{\mathfrak{q}}_\pm(\lambda): \lambda\in \mathrm{spec}(J)\}. \]
All the real parts of the eigenvalues of $\mathcal{H}^{(R)}$ are all the eigenvalues of $J$. On the other hand, 
the rest of eigenvalues are given by $1$ and $-1$ with the multiplicities $|E|-|V|-|\mathcal{S}|+m(1)$ and $|E|-|V|+m(-1)$, respectively. 
\end{remark}
In the graph without both cycles nor self loops, then the dimension of our quantum walk is reduced to $d=2|V|-2$. 
It is known that whe the graph is a path, then the spectrum of $J$, $\mathrm{spec}(J)$, has eigenvalue $1$ without multiplicity. 
If $-1\notin \mathrm{spec}(J)$, then from the above lemma, $d=2|V|-1$ which contradicts to the dimension of the walk. 
Thus $-1\in \mathrm{spec}(J)$. 
\begin{remark}\label{tama}
If the graph is a path, then $\mathcal{H}^{(S)}=\emptyset$. 
\end{remark}
Let $\boldsymbol{P}(x)$ satisfy the following eigenequation. 
\[ M\boldsymbol{P}(x)=x\boldsymbol{P}(x). \] 
The reversible Markov chain implies $\boldsymbol{\mathfrak{p}}(x)=D^{-1/2}\boldsymbol{P}(x)$ which provides following remark: 
\begin{remark}\label{katsumori2}
If the Markov chain is reversible whose limit measure $\bs{\pi}(u)> 0$ for all $u\in V(G)$, then we can reexpress the eigenvectors of $U$ as 
\[\boldsymbol{\mathfrak{q}}_{\pm}(x)
	=\begin{cases}
        D^{-1/2}(I-e^{\pm i\theta(x)}S)A\bs{P}(x) & \text{: $x\notin \{\pm 1\}$, } \\
        D^{-1/2}A\bs{P}(x) & \text{: $x\in \{\pm 1\}$.}
         \end{cases} 
        \]
\end{remark}
\noindent \\
\noindent Now, we give the proof of Theorem 1. \\
\noindent {\it {Proof.}} 
It should be noticed that $AA^\dagger=\Pi_\mathcal{A}$ , $BB^\dagger=\Pi_\mathcal{B}$. 
Then we have
\[ A\boldsymbol{\mathfrak{p}}(x) \stackrel{C}{\mapsto} A\boldsymbol{\mathfrak{p}}(x) \stackrel{S}{\mapsto} B\boldsymbol{\mathfrak{p}}(x). \]
On the other hand, from $A^\dagger B=B^\dagger A=J$ and the definition of $\boldsymbol{\mathfrak{p}}(x)$, 
\[  B\boldsymbol{\mathfrak{p}}(x)\stackrel{C}{\mapsto} 2xA\boldsymbol{\mathfrak{p}}(x)-B\boldsymbol{\mathfrak{p}}(x)
	\stackrel{S}{\mapsto} 2xB\boldsymbol{\mathfrak{p}}(x)-A\boldsymbol{\mathfrak{p}}(x). \]
Therefore the subspace spanned by $A\boldsymbol{\mathfrak{p}}(x)$ and $B\boldsymbol{\mathfrak{p}}(x)$ is invariant under the action of $U$. 
Thus we arrive at the conclusion of part (1) of the lemma. \\
It is noticed that $\boldsymbol{s}\in \mathcal{H}^{(S)}$ implies $S\boldsymbol{s}\in \mathcal{H}^{(S)}$. 
We find that
\[ \boldsymbol{s} \stackrel{C}{\mapsto} -\boldsymbol{s} \stackrel{S}{\mapsto} -S\boldsymbol{s}, \;\;\; 
S\boldsymbol{s} \stackrel{C}{\mapsto} -S\boldsymbol{s} \stackrel{S}{\mapsto} -\boldsymbol{s}. \]
Therefore $U(I\pm S)\boldsymbol{s}=\mp (I\pm S)\boldsymbol{s}$, for all $\boldsymbol{s}\in \mathcal{H}^{(S)}$. 
Since for any $\boldsymbol{s}\in \mathcal{H}^{(S)}$, $((I+S)\boldsymbol{s}+(I-S)\boldsymbol{s})/2=\boldsymbol{s}$, 
\begin{equation}\label{div} 
\mathcal{H}_+=\mathrm{span}\{ (I+S)\boldsymbol{s}; \boldsymbol{s}\in \mathcal{H}^{(S)} \},\; \mathcal{H}_-=\mathrm{span}\{ (I-S)\boldsymbol{s}; \boldsymbol{s}\in \mathcal{H}^{(S)} \}. 
\end{equation}
Note that 
\begin{equation}\label{s_eigen}
S(I+ S)\bs{s}=(I+ S)\bs{s},\;\;S(I-S)\bs{s}=-(I-S)\bs{s}. 
\end{equation}
Equations (\ref{div}) and (\ref{s_eigen}) implies that 
\begin{equation}
\mathcal{H}_+=\mathcal{H}^{(S)}\cap \mathcal{M}_-,\;\;\mathcal{H}_-=\mathcal{H}^{(S)}\cap \mathcal{M}_+. 
\end{equation}
Since 
\begin{align*}
S\left(A\boldsymbol{\mathfrak{p}}(x)+ B\boldsymbol{\mathfrak{p}}(x)\right) &= \left(A\boldsymbol{\mathfrak{p}}(x)\pm B\boldsymbol{\mathfrak{p}}(x)\right), \\
S\left(A\boldsymbol{\mathfrak{p}}(x)- B\boldsymbol{\mathfrak{p}}(x)\right) &= -\left(A\boldsymbol{\mathfrak{p}}(x)- B\boldsymbol{\mathfrak{p}}(x)\right),\;\;\mathrm{for\;} x\in \mathrm{spec}(J)\setminus{\{\pm 1\}}, \\
SA\mathfrak{p}(1) &= A\mathfrak{p}(1),\;\;\mathrm{for\;} 1\in \mathrm{spec}(J), \\
SA\mathfrak{p}(-1) &= -A\mathfrak{p}(-1),\;\;\mathrm{for\;}-1\in \mathrm{spec}(J), 
\end{align*}
it is hold that 
\[ \mathrm{dim}\left( \mathcal{H}^{(R)}\cap \mathcal{M}_+ \right)=|V|-m(-1),\;\;\mathrm{dim}\left( \mathcal{H}^{(R)}\cap \mathcal{M}_- \right)=|V|-m(1). \]
Combining Remark~\ref{shiftoperator} with it, we obtain 
\begin{align*}
\mathrm{dim}(\mathcal{H}^{(S)}_+) &= \mathrm{dim}(\mathcal{H}^{(S)}\cap \mathcal{M}_- )=\mathrm{dim}(\mathcal{M}_-)-\mathrm{dim}(\mathcal{H}^{(R)}\cap \mathcal{M}_-) \\
	&= |E|-|S|-(|V|-m(1)). 
\end{align*}
In the same way, 
\[ \mathrm{dim}(\mathcal{H}^{(S)}_-) = |E|-(|V|-m(-1)). \]
Then we get the desired conclusion of part (2) of the lemma. 
\begin{flushright} $\square$ \end{flushright}
\section{Sufficient conditions of localizations given by recurrence properties of RWs}
\subsection{Time averaged limit measure of QW}
The set of time averaged limit measures $\overline{\mathcal{M}}_\infty$ 
is defined as follows: for any $u,v\in V(G)$, 
\[ \overline{\mathcal{M}}_\infty (u,v)=\left\{\lim_{T\to\infty}\frac{1}{T}\sum_{t=0}^{T-1}P(X_t^{(\Psi_0)}=u): 
	\bs{\Psi}_0\in \left\{ \sum_{w:(w,v)\in A(G)}\alpha_{w}\bs{\delta}_{(w,v)}: \sum_{w:(w,v)\in A(G)}|\alpha_w|^2=1 \right\} \right\}. \]
This measure is so called Ces{\'a}ro summation of the finding probability from $v$ with the initial state $\bs{\Psi}_0$ to $u$. 
\begin{lemma}\label{acco2} 
For any $(u,v)\in V\times V$, the time averaged limit measure 
$\overline{\mu}_\infty^{(\Psi_0)}(u,v)\in \overline{\mathcal{M}}_\infty(u,v)$ with the initial state 
$\bs{\Psi}_0=\sum_{w:(w,v)}\alpha_w \bs{\delta}_{(w,v)}$ satisfies the following statements.  
If the corresponding Markov chain has the reversible distribution $\bs{\pi}$, 
then the time averaged limit measure 
$\overline{\mu}_\infty^{(\Psi_0)}(u,v)$ is bounded below as follows:  
\begin{equation}
\overline{\mu}_\infty^{(\Psi_0)}(u,v)\geq  |\langle \bs{a}_v,\bs{\Psi}_0 \rangle|^2 \bs{\pi}(u) \bs{\pi}(v)
	+\sum_{w:(w,u)\in A}|\langle \bs{\delta}_{(w,u)},\Pi_{\mathcal{S}}\bs{\Psi}_0 \rangle|^2,
\end{equation}
where $\Pi_{\mathcal{S}}$ is the orthogonal projection onto $\mathcal{H}^{(\mathcal{S})}$. 
\end{lemma}
\noindent {\it {Proof.}} 
We consider the QW on $G(V,E)$ with $|V|=\infty$. 
The time evolution $U$ is induced by the Jacobi matrix $J$. 
Let $\bs{\mathfrak{p}}(x)$ and $\bs{\mathfrak{q}}_\pm (x)$ satisfy the eigenequations $J\bs{\mathfrak{p}}(x)=x\bs{\mathfrak{p}}(x)$ and 
$U\bs{\mathfrak{q}}_\pm(x)=x\bs{\mathfrak{q}}_\pm(x)$, respectively. 
Fix a vertex $v_0\in V$ called the origin and 
let us regard $G$ as the graph $G_n=G_n(V_n,E_n)$ in the limit of $n\to \infty$, where 
$V_n=\{ v\in V: \rho(v_0,v)\leq n \}$, and $E_n=\{\{u,v\}\in E : u,v\in V_n \}$. Here $\rho(u,v)$ is the shortest distance between $u$ and $v$. 
Put the time evolution of the QW on $G_n$ as $U_n$, and $J_n$ as the corresponding Jacobi matrix. 
Assume that the time evolution $U_n$ is decomposed into 
\[ U_n=\sum_{k}e^{i\theta_k^{(n)}}\bs{v}_k^{(n)}{\bs{v}_k^{(n)}}^{\dagger}, \]
where $\bs{v}_k^{(n)}$ is the normalized eigenvector of eigenvalue $e^{i\theta_k^{(n)}}$. 
Lemma \ref{acco1} and $||\bs{\mathfrak{q}}_{\pm}^{(n)}(\lambda)||^2=2(1-\lambda^2)$  ($\lambda\in \mathrm{spec}(J_n)\setminus \{\pm 1\}$), 
$=1$ ($\lambda \in \mathrm{spec}(J_n)\cap \{\pm 1\}$) provide
\[ \{\bs{v}_k^{(n)}\}_k\supseteq 
	\left\{ \left( \frac{1-\delta_{\{|\lambda|=1\}}}{\sqrt{2(1-\lambda^2)}}+\delta_{\{|\lambda|=1\}}\right) \bs{\mathfrak{q}}_\pm^{(n)}(\lambda): 
        \lambda \in \mathrm{spec}(J_n)\right\}. \] 
We should remark that $\lim_{T\to\infty}\sum_{t=0}^{T-1}e^{it(\theta_l^{(n)}-\theta_m^{(n)})}/T=\delta_{l,m}$. 
The time averaged limit measure $\overline{\mu}_\infty^{(\Psi_0,n)}(u,v)$ with the initial state 
$\bs{\Psi}_0=\sum_{w:(w,v)\in A_n}\alpha_w \bs{\delta}_{(w,v)}$ is reexpressed by 
\begin{equation} 
\overline{\mu}_\infty^{(\Psi_0,n)}(u,v)
	=\sum_{k}\sum_{w: (w,u)\in A_n}|\langle \bs{v}_k^{(n)}, \Psi_0\rangle |^2 |\langle \bs{v}_k^{(n)}, \bs{\delta}_{(w,u)}\rangle|^2. 
\end{equation}
In the limit of $n\to\infty$, put 
\begin{equation}\label{B}
\mathcal{B}=\{ x_0 \in [-1,1]: \bs{\mathfrak{p}}(x_0)\in \ell^2(V)\setminus\{0\} \}
\end{equation}
as the set of the mass points of $J$. 
Equation (\ref{pero1}) in Lemma \ref{acco1} implies that 
if $\bs{\mathfrak{p}}(x_0)\in \ell^2(V)$, then $\bs{\mathfrak{q}}_{\pm}(x_0)\in \ell^2(A)$. 
Let $\overline{\mu}_\infty^{(\Psi_0)}(u,v)=\lim_{n\to\infty}\overline{\mu}_\infty^{(\Psi_0,n)}(u,v)$. 
Then we have 
\begin{multline}\label{infinite_TALM}
\overline{\mu}_\infty^{(\Psi_0)}(u,v)
	= \sum_{x_0\in \mathcal{B},\;\epsilon\in\{\pm\}}\sum_{w:(w,u)\in A}
        \left( \frac{1-\delta_{\{|x_0|=1\}}}{4(1-x_0^2)^2}+\delta_{\{|x_0|=1\}}\right)
        \left|\langle \bs{\mathfrak{q}}_\epsilon(x_0), \Psi_0\rangle \right|^2 \left|\langle \bs{\mathfrak{q}}_\epsilon(x_0), \bs{\delta}_{(w,u)}\rangle\right|^2 \\
        +\sum_{w:(w,u)\in A}|\langle \bs{\delta}_{(w,u)},\Pi_{\mathcal{S}}\bs{\Psi}_0 \rangle|^2. 
\end{multline}
Since the corresponding Markov chain has a reversible distribution, combining Lemma \ref{acco1} and Remark \ref{acco2} with Eq. (\ref{infinite_TALM}) implies 
\begin{align} 
\overline{\mu}_\infty^{(\Psi_0)}(u,v)
	\geq |\langle \bs{a}_v,\bs{\Psi}_0 \rangle|^2 \bs{\pi}(u) \bs{\pi}(v)
	+\sum_{w:(w,u)\in A}|\langle \bs{\delta}_{(w,u)},\Pi_{\mathcal{S}}\bs{\Psi}_0 \rangle|^2. 
\end{align}
\begin{flushright}$\square$\end{flushright}

\begin{remark}
If $|V|<\infty$, then the time averaged limit measure is a distribution in that $\sum_{u\in V}\overline{\mu}_\infty^{(\Psi_0)}(u,v)=1$. 
On the other hand, if $|V|=\infty$, then its distributivity is not ensured in that $\sum_{u\in V}\overline{\mu}_\infty^{(\Psi_0)}(u,v)\leq 1$
as is seen in \cite{IKS}, for example. 
\end{remark}
\subsection{Localization of QW on half line}
In this subsection, we treat a QW on a simple graph, semi infinite line with some self loops. 
We say localization happens at position $i\in \{0,1,2,\dots\}$ when the time averaged limit measure 
with the initial state $\bs{\Psi}_0=\sum_{k:|k-j|\leq 1}\alpha_{k,j}\bs{\delta}_{(k,j)}$ satisfies 
\[ \overline{\mu}_\infty^{(\Psi_0)}(i,j)>0. \]
The stochastic matrix of the RW which induces the corresponding QW is given by 
\[
M=
\left(
\begin{array}{ccccccc}
 r_0 & q_1 & 0   &        &        &  &  \\
 p_0 & r_1 & q_2 & 0      &        &  &  \\
 0   & p_1 & r_2 & q_3    & 0      &  &  \\
     & 0   & p_2 & \ddots & \ddots &  &  \\
     &     &     & \ddots &        &  &  \\
\end{array}
\right),
\]
where $q_j+r_j+p_j=1$, $p_j>0$ $(j\geq 0)$, $q_j>0$ $(j\geq 1)$. 
A particle at position $j$ jumps to left, right and itself with probabilities $q_j$, $p_j$ and $r_j$, respectively at each time, independently. 
Thus the symmetric oriented graph $G'(V,A)$ treated in this subsection is $V(G')=\{0,1,2,\dots\}$, $A(G')=\{(i,j):|i-j|=1\}\cup \{(i,i):r_i\neq 0 \}$. 
At first, we review the recurrence properties of the RW on a half line. For more details, see \cite{Schinazi}, for example. 
Define 
\[ C_T=\sum_{j\geq 1}\frac{q_1\cdots q_j}{p_1\cdots p_j},\;\;\;\;C_R=\sum_{j\geq 1}\frac{p_0\cdots p_{j-1}}{q_1\cdots q_j}. \]
If $C_T=\infty$, then the RW is transient, otherwise recurrent. Moreover if $C_R=\infty$ in the recurrent RW, then 
the RW is null recurrent, otherwise positive recurrent. See also Table \ref{table1}. 
\begin{table}
\begin{center}
\begin{tabular}{|r|c|c|c|}
\hline
                       & Transient & Null recurrent & Positive recurrent \\ \hline
Recurrence probability & $<1$      & \multicolumn{2}{|c|}{$=1$}          \\ \hline
$C_T$                  & $<\infty$ & \multicolumn{2}{|c|}{$=\infty$}     \\ \hline
Mean recurrence time   & \multicolumn{2}{|c|}{$=\infty$} & $<\infty$       \\ \hline
$C_R$                  & \multicolumn{2}{|c|}{$=\infty$} & $<\infty$       \\ \hline
\end{tabular}
\end{center}
\caption{Recurrence properties of RW}
\label{table1}
\end{table}
We see the following interesting example given by \cite{Schinazi}: 
In the space homogeneous case $p_j=p$, and $q_j=q$  ($j\geq 1$), then the RW is transient if $p>q$, 
if $p=q$, null recurrent, and if $p<q$, positive recurrent. 
On the other hand, in the following three cases (a), (b) and (c), we see that $p_\infty=q_\infty$ in the limit of $i\to \infty$. 
However, we find that the three cases have different recurrence properties: 
case (a) is transient, while cases (b) and (c) are null and positive recurrent, respectively. 
\begin{enumerate}
\renewcommand{\labelenumi}{(\alph{enumi})}
\item $p_i=(i+2)/(2i+2)$, $q_i=i/(2i+2)$, 
\item $p_i=(i+1)/(2i+1)$, $q_i=i/(2i+1)$, 
\item $p_0=1$, $p_1=q_1=1/2$, $p_i=(i-1)/2i$, $q_i=(i+1)/2i$, $(i\geq 2)$. 
\end{enumerate}
Even if we add some finite number of self loops $\mathcal{S}=\{j_1,j_2,\dots,j_n\}\subset \mathbb{Z}_+$ to the above RWs so that 
$p_j+q_j<1$ for every $j\in \mathcal{S}$, there is no influence to these recurrence properties. 
This is a trivial statement but important to this section. 
From now on, 
we give the following theorem with respect to localization of the QW induced by recurrence properties of the corresponding RW. 
\begin{theorem}\label{machiko}
If the RW is positive recurrent with the stationary distribution $\bs{\pi}$, then an appropriate choice of the initial state 
gives localization of the corresponding QW. 
Moreover, its time averaged limit measure with the initial state $\bs{\Psi}_0=\sum_{k:|k-j|\leq 1}\alpha_{k}\bs{\delta}_{(k,j)}$ 
is bounded below as follows.
\begin{equation}
\overline{\bs{\mu}}_\infty^{(\Psi_0)}(i,j)\geq |\langle \bs{a}_j,\bs{\Psi}_0\rangle|^2\bs{\pi}(i)\bs{\pi}(j).
\end{equation}
Here
\begin{equation}\label{SD} \bs{\pi}(j)=\frac{1}{1+C_R}\left\{\delta_0(j)+(1-\delta_0(j))\frac{p_0\cdots p_{j-1}}{q_1\cdots q_j}\right\}. \end{equation}
\end{theorem}
\noindent {\it {Proof.}} 
The corresponding Jacobi matrix is given by 
\[
J=
\left(
\begin{array}{ccccccc}
 r_0 & \sqrt{p_0q_1} & 0 &  &  &  &  \\
 \sqrt{p_0q_1} & r_1 & \sqrt{p_1q_2} & 0 &  &  &  \\
 0 & \sqrt{p_1q_2} & r_2 & \sqrt{p_2q_3} & 0 &  &  \\
  & 0&  \sqrt{p_2q_3} & \ddots & \ddots &  &  \\
  &  &  & \ddots &  &  &    
\end{array}
\right).
\]
The detailed balance condition $\pi_ip_i=\pi_{i+1}q_{i+1}$ for $i\geq 0$ and positive recurrence of the RW 
provide $J=D^{1/2}MD^{-1/2}$ and Eq. (\ref{SD}), 
where $D$ is a diagonal matrix with $(D)_{i.i}=\bs{\pi}(i)$.
Therefore from Lemma \ref{acco2}, we obtain the desired conclusion. 
\begin{flushright}$\square$\end{flushright}
\begin{remark}
If we want to avoid localization of the QW launched at $j$ induced by the positive recurrent RW, at least, we should choose the three parameters of initial state 
$\alpha_{j-1}, \alpha_{j}, \alpha_{j+1}$ 
so that $\bs{\varphi}_0\equiv {}^T[\alpha_{j-1}, \alpha_{j}, \alpha_{j+1}]$ is orthogonal to ${}^T[\sqrt{q_j},\sqrt{r_j},\sqrt{p_j}]$.
\end{remark}
Let us consider a special case $r_j=0$ for all $j\in \mathbb{Z}_+$ in the following. 
In this case, from Remark \ref{tama}, it is obtained that $\mathcal{H}^S=\emptyset$. 
Now we check whether $-1$ is eigenvalue of $U$ or not. 
In a heuristic argument, a ``signed"-detailed balance condition $\bs{\pi}'(j)p_j=-\bs{\pi}'(j+1)q_{j+1}$ implies that 
$\bs{\pi}'(j)=(-1)^{j}(p_{j-1}\cdots p_{0})/(q_j\cdots q_{1})$. 
Actually, we see 
\begin{equation}\label{signed}  M\bs{\pi}'=-\bs{\pi}'. \end{equation}
Recall that $\bs{\mathfrak{p}}(x)$ is the eigenvector of $J$ with its eigenvalue $x$ defined in Eq. (\ref{yoshio}). 
Combining Eq. (\ref{signed}) with Remark \ref{katsumori2} gives $\bs{\mathfrak{p}}_j(-1)=(-1)^j\bs{\mathfrak{p}}_j(1)$ 
which implies that $U$ has the eigenvalue $-1$ from Eq. (\ref{pero1}) in Lemma \ref{acco1}. 
Also we know from Eq. (\ref{masako}) in Lemma \ref{acco1} that one of the derivations of the value $-1$ is the eigenvalue of $\mathcal{H}^{(S)}$, 
but in this case $\mathcal{H}^{(S)}=\emptyset$. 
So we find that the derivation of the eigenvalue $-1$ comes from just only $\mathcal{H}^{(R)}$ in this case. 
Then from Eq. (\ref{infinite_TALM}), we obtain the following corollary. 
\begin{corollary}
Let $r_j=0$ for all $j\in \mathbb{Z}_+$. If the RW is positive recurrent, then the time averaged limit measure is bounded below as follows
\begin{equation}\label{IKS}
\overline{\bs{\mu}}_\infty^{(\Psi_0)}(i,j)\geq 2\times |\langle \bs{a}_j,\bs{\Psi}_0\rangle|^2\bs{\pi}(i)\bs{\pi}(j), 
\end{equation}
where $\bs{\pi}$ is the stationary distribution of the RW given in Eq. (\ref{SD}). 
\end{corollary}
\begin{remark}
Assume that we choose the initial coin state from one of the following set: for $j\geq 1$ 
\begin{equation}
\bs{\Psi}_0=
	\begin{cases}
        \bs{\delta}_{(j+1,j)} & \text{: with prob. $1/2$,} \\
        \bs{\delta}_{(j-1,j)} & \text{: with prob. $1/2$,}
        \end{cases}
\end{equation}
for $j=0$, $\bs{\Psi}_0=|0;R\rangle$. 
Let us consider the space homogeneous case such that $p_j=p$ $(j\geq 1)$, $=1$ $(j=0)$, $q_j=q$ $(j\geq 1)$, and $r_j=0$ $(j\geq 0)$ with $q>p$ (positive recurrent). 
Then the description of section 4 in \cite{Grun} with respect to the presence of two point masses $\{\pm 1\}$ of the 
spectral measure of a Chebyshev type stochastic matrix, ensures that 
the time averaged limit measure is described by the lower bound itself in Eq. (\ref{IKS}), that is, 
\begin{align}
\overline{\bs{\mu}}_\infty^{(\Psi_0)}(i,j)
	&=\big\{2\delta_0(j)+\left(1-\delta_0(j)\right)\big\}\bs{\pi}(j)\bs{\pi}(i), 
\end{align}
where $\bs{\pi}(j)=(1-p/q)/2$ $(j=0)$, $=(1-p/q)/(2q)\times (p/q)^{j-1}$ $(j\geq 1)$. 
Thus the summation of the time averaged limit measure $c_j$ is given by 
\[ c_j=\big\{ 2\delta_0(j)+(1-\delta_0(j)) \big\}\bs{\pi}(j)<1, \]
which is consistent with the argument of \cite{IdeKS}, which treats a space homogeneous QW on a finite path, in the limit of its system size. 
\end{remark}
\noindent \\
Localization of Theorem \ref{machiko} and its associated arguments come from $\mathcal{H}^{(R)}\subset \mathcal{H}$. 
Recall that the derivation of $\mathcal{H}^{(R)}$ is the corresponding RW. 
Since the subspace $\mathcal{H}^{(R)}$ is invariant under the action of $U$, 
the choice of initial state from $\mathcal{H}^{(R)}$ leads us considerations outside of its orthogonal complement $\mathcal{H}^{(S)}$. 
From now on, we will face the consideration of $\mathcal{H}^{(S)}$. 
For a simplicity, we change the notations of the basis of $\mathcal{H}$ as follows: 
$\bs{\delta}_{(j+1,j)}\to |j;R\rangle$, $\bs{\delta}_{(j-1,j)}\to |j;L\rangle$, and if $r_j\neq 0$, then 
$\bs{\delta}_{(j,j)}\to |j;O\rangle$. 
\begin{lemma}\label{akio}
Assume that $\mathcal{S}=\{ j_1,j_2,\dots \}$, $(j_1<j_2<\cdots)$. Put 
\[ Q_j=
	\begin{cases}  
	1 & \text{: $j=0$,} \\
        \sqrt{\frac{q_1\cdots q_{j}}{p_1\cdots p_jq_j}} & \text{: $j\geq 1$.} 
	\end{cases} 
\]
Then it is obtained that 
$\mathcal{H}^{(S)}$ is spanned by $\{\bs{\eta}_k\}_{j_k\in \mathcal{S} }$, 
where
\begin{multline}
\bs{\eta}_k=
(-1)^{j_k}{Q}_{j_k}\left(-\sqrt{\frac{p_{j_k}\tilde{q}_{j_k}}{r_{j_k}}}|j_k;O\rangle+\sqrt{\tilde{q}_{j_k}}|j_k;R\rangle \right)
+I_{\{j_{k+1}-j_k>1\}}\sum_{l=j_k+1}^{j_{k+1}-1}(-1)^{l} {Q}_l\bigg(-\sqrt{p_l}|l;L\rangle+ \sqrt{q_l}|l;R\rangle\bigg) \\
+(-1)^{j_{k+1}}{Q}_{j_{k+1}}\left(-\sqrt{p_{j_{k+1}}}|j_{k+1};L\rangle+\sqrt{\frac{p_{j_{k+1}}q_{j_{k+1}}}{r_{j_{k+1}}}}|j_{k+1};O\rangle \right).
\end{multline}
Here $\tilde{q}_j=q_j$ $(j\geq 1)$, $=1$ $(j=0)$.
In particular, if $|\mathcal{S}|=n<\infty$, then 
\begin{equation}\label{makiko}
\bs{\eta}_n=
(-1)^{j_{n}}{Q}_{j_n}\left(-\sqrt{\frac{p_{j_n}q_{j_n}}{r_{j_n}}}|j_n;O\rangle+\sqrt{q_{j_n}}|j_n;R\rangle \right)
+\sum_{l=j_n+1}^{\infty}(-1)^{l} {Q}_l\bigg(-\sqrt{p_l}|l;L\rangle+ \sqrt{q_l}|l;R\rangle\bigg).
\end{equation}
\end{lemma}
\noindent \\
\noindent \\
In this paper, we call $\bs{\eta}_k$ ``signed reflected vector" since $U\bs{\eta}_k=-\bs{\eta}_k$. Now we give the proof of Lemma \ref{akio}. \\
\noindent \\
\noindent \\
\noindent {\it {\large proof.}} 
Remark that $\mathcal{H}^{(R)}=\mathrm{span}\{\bs{a}_j,\bs{b}_j: j\geq 0\}$, where $\bs{a}_j=\sqrt{q_j}|j;L\rangle+\sqrt{r_j}|j;O\rangle+\sqrt{p_j}|j;R\rangle$, and 
$\bs{b}_j=S\bs{a}_j=\sqrt{q_j}|j-1;R\rangle+\sqrt{r_j}|j;O\rangle+\sqrt{p_j}|j+1;L\rangle$. 
Put $\bs{\phi}\in \mathcal{H}^{(S)}$ as 
\begin{equation}\label{hinano3}
\bs{\phi}=\sum_{j\geq 0}\alpha^{(-)}_j|j;L\rangle+\alpha^{(o)}_j|j;O\rangle+\alpha^{(+)}_j|j;R\rangle,
\end{equation}
where 
if $r_j=0$, then $\alpha_j^{(o)}=0$. 
Then $\langle \bs{\phi}, \bs{a}_j \rangle=\langle \bs{\phi}, \bs{b}_j \rangle=0$ for any $j$ implies 
\begin{align}
\sqrt{q_j}\alpha_j^{(-)}+\sqrt{r_j}\alpha_j^{(o)}+\sqrt{p_j}\alpha_j^{(+)} &= 0, \label{hinano1}\\
\sqrt{q_j}\alpha_{j-1}^{(+)}+\sqrt{r_j}\alpha_j^{(o)}+\sqrt{p_j}\alpha_{j+1}^{(-)} &= 0. \label{hinano2}
\end{align}
The above equations lead to 
\[ \alpha_{j+1}^{(-)}-\alpha_{j}^{(+)}=\sqrt{\frac{q_j}{p_j}}\left(\alpha_{j}^{(-)}-\alpha_{j-1}^{(+)}\right)
	=\sqrt{\frac{q_j\cdots q_1}{p_j \cdots p_1}}\left(\alpha_1^{(-)}-\alpha_0^{(+)}\right). \]
On the other hand, from the $j=0$ cases in Eqs. (\ref{hinano1}), (\ref{hinano2}), we have $\alpha_1^{(-)}=\alpha_0^{(+)}$. 
After all, we get $\alpha_j \equiv \alpha_{j+1}^{(-)}=\alpha_{j}^{(+)}$ for $j\geq 0$. 
Here we should note that 
if $r_j=0$, then $\alpha_j^{(o)}=0$ and $\alpha_j=-\sqrt{q_j/p_j}\alpha_{j-1}$, otherwise, $\alpha_j^{(o)}=-\sqrt{q_j/r_j}\alpha_{j-1}-\sqrt{p_j/r_j}\alpha_{j}$. 
Substituting these relations into RHS of Eq. (\ref{hinano3}), we arrive at the conclusion.  
\begin{flushright}$\square$\end{flushright}
For any subspace $\mathcal{H}'\subset \mathcal{H}$, define 
\[\mathrm{supp}(\mathcal{H}')=\left\{j\in \mathbb{Z}_+: {}^\exists \bs{\phi}\in \mathcal{H}'\;\mathrm{and}\;{}^\exists J\in\{R,O,L\}, \mathrm{s.t.}\; 
\langle \bs{\phi}|j;J\rangle \neq 0 \right\},
\] 
where $\mathbb{Z}_+=\{0,1,2,\dots\}$. 
From Eq. (\ref{infinite_TALM}), when we choose the initial state $\bs{\Psi}_0$ so that 
$\bs{\Psi}_0\in \mathrm{span}\{\bs{\mathfrak{q}}_\pm (x_0): x_0\in \mathcal{B}\}^{\bot}$, 
then the contribution of localization of the QW is only $||\Pi_S \bs{\Psi}_0||^2$, where $\mathcal{B}$ is defined in Eq. (\ref{B}). 
We should remark that the support of localization is included by $\mathrm{supp}(\mathcal{H^{(S)}})$, 
that is $\mathrm{supp}(\{ \Pi_S \bs{\Psi}_0 \})\subseteq \mathrm{supp}(\mathcal{H^{(S)}})$. 
 
The subspace $\mathcal{H}^{(S)}$ is outside of $\mathcal{H}^{(R)}$ which is 
induced by the corresponding RW. 
However the recurrence properties of the RW appear again roundabout 
to the following theorem with respect to $\mathrm{supp}(\mathcal{H^{(S)}})$. (see also Table 2.)
\begin{theorem}
Suppose that the set of self loops is denoted by $\mathcal{S}=\{j_1,j_2,\dots\}$, $(j_1<j_2<\cdots)$ with $j_{k+1}-j_k<\infty$. 
Then it is obtained that 
\begin{enumerate}
\item $|\mathcal{S}|=0$ case. $\mathrm{supp}(\mathcal{H}^{(S)})=\emptyset$. 
\item $1\leq |\mathcal{S}|=n<\infty$ case. 
\[ \mathrm{supp}(\mathcal{H}^{(S)})=
   \begin{cases}
   \{j\in \mathbb{Z}_+: j_1\leq j\} & \text{if the RW is transient,} \\
   \emptyset & \text{if the RW is recurrent and $n=1$, } \\
   \{j\in \mathbb{Z}_+: j_1 \leq j\leq j_n\} & \text{if the RW is recurrent and $n>1$. } \\
   \end{cases} 
\]
\item $|\mathcal{S}|=\infty$ case. 
\[ \mathrm{supp}(\mathcal{H}^{(S)})=\{j\in \mathbb{Z}_+: j_1\leq j\}. \]
\end{enumerate}
\end{theorem}
\begin{table}
\begin{center}
\begin{tabular}{|c|c|c|}
\hline
 & recurrent & transient \\ \hline
$|\mathcal{S}|=0$ & \multicolumn{2}{|c|}{$\emptyset$} \\ \hline
$|\mathcal{S}|=1$ & $\emptyset$ & $\{j_1\leq j\}$ \\ \hline
$1<|\mathcal{S}|\leq n$ & $\{j_1\leq j\leq j_n\}$ & $\{j_1\leq j\}$ \\ \hline
$|\mathcal{S}|=\infty$ & \multicolumn{2}{|c|}{$\{j_1\leq j\}$} \\ \hline
\end{tabular}
\caption{$\mathrm{supp}(\mathcal{H}^{(S)})$}
\end{center}
\end{table}
\noindent {\it {Proof.}} 
If $|\mathcal{S}|=0$, then the graph becomes a tree structure. Then from Remark \ref{tama}, $\mathcal{H}^{(S)}=\emptyset$ which is the conclusion of part (1) of the Theorem 2. 
Next we consider the case $|\mathcal{S}|=\infty$. 
From Lemma \ref{akio}, all the signed reflected vectors $||\bs{\eta}_k||<\infty$ $(k\geq 0)$, that is, $\bs{\eta}_k\in \ell^2(A)$ 
since $\mathrm{supp}(\{\eta_k\})=j_{k+1}-j_k<\infty$. Thus we get the conclusion of part (3) of Theorem 2. 
Finally, we consider the case $0<|\mathcal{S}|=n<\infty$. In this case, the assumption $j_{k+1}-j_k<\infty$ ensures 
for all $0\leq j\leq n-1$, $||\bs{\eta}_j||^2<\infty$. 
But the $\ell^2$ summability of the final signed reflected vector $\bs{\eta}_n$ is not ensured by the assumption 
because the support of $\bs{\eta}_n$ is $\{j\geq n\}$. Now we should check the value of $||\bs{\eta}_n||^2$ in the following. 
Equation (\ref{makiko}) in Lemma \ref{akio} provides
\begin{align}\label{pipi}
||\bs{\eta}_n||^2 &= \frac{q_{j_n}(1-q_{j_n})}{r_{j_n}}Q_{j_n}^2+\sum_{l=j_n+1}^\infty Q_l^2. 
\end{align} 
Note $p_l+q_l=1$ for $l > j_n$, since $j_n$ the right most self loop. By using this fact, we get 
\begin{align} 
\sum_{l= j_n+1}^\infty Q_l^2 
	&= \sum_{l= j_n+1}^\infty \frac{q_1\cdots q_{l-1}}{p_1\cdots p_{l}}= \sum_{l=j_n+1}^\infty \frac{q_1\cdots q_{l}}{p_1\cdots p_{l}}\frac{p_l+q_l}{q_l} \notag \\
        &= 2\sum_{l= j_n+1}^\infty \frac{q_1\cdots q_{l}}{p_1\cdots p_{l}}
        +\left\{ (1-I_{\{j_n=0\}})\frac{q_1\cdots q_{j_n}}{p_1\cdots p_{j_n}}+I_{\{j_n=0\}}\frac{1-q_1}{p_1}\right\}. \label{pooh}
\end{align}
When the RW is transient, then $\sum_{l= j_n+1}^\infty (q_1\cdots q_{l})/(p_1\cdots p_{l})\leq C_T<\infty$. Combining Eqs. (\ref{pipi}) with (\ref{pooh}), 
we obtain the desired conclusion of part (2). Then we complete the proof. 
\begin{flushright}$\square$\end{flushright}
In the following two corollaries, to highlight effects of $\mathcal{H}^{(S)}$ to localization, 
we impose the following assumptions to the initial state $\bs{\Psi}_0$: 
\begin{enumerate}
\renewcommand{\labelenumi}{(\theenumi)}
\item $\bs{\Psi}_0\in \mathrm{span}\{\bs{\mathfrak{q}}_\pm (x_0): x_0\in \mathcal{B}\}^{\bot}$,
\item $\mathrm{supp}(\bs{\Psi}_0)<\infty$. 
\end{enumerate}
Define $\overline{\mu}_\infty^{(\Psi_0)}(j)=\lim_{T\to\infty}1/T \sum_{t=0}^{T-1}P(X_t^{(\Psi_0)}=j)$. 
We give simple examples of Theorem 2 as the following corollaries which give time averaged limit measures by directly computing 
$\sum_{J\in\{L,O,R\}}|\langle i;J| \Pi_\mathcal{S}\bs{\Psi}_0\rangle|^2$. 
\begin{corollary}
Suppose that $\mathcal{S}=\{0\}$. 
If the RW is transient, 
then the time averaged limit measure of the corresponding QW with the initial state 
$\bs{\Psi}_0$ satisfying the assumptions (1) and (2) 
is given by 
\begin{equation}\label{kataken}
\overline{\mu}_\infty^{(\Psi_0)}(i) = \left|\sum_{j\geq 0} \langle \bs{a}^{\bot}_j, \bs{\Psi}_0 \rangle \right|^2\bs{\pi'}(i)
\end{equation}
where 
\begin{align} 
\bs{\pi'}(j) &= \frac{1}{1+C_R'}\left\{ \delta_{0}(j)+\left(1-\delta_{0}(j)\right)\frac{r_0q_1\cdots q_{j-1}}{p_1\cdots p_j}  \right\}, \\
\bs{a}^{\bot}_j &= (-1)^j\sqrt{\bs{\pi'}(j)}\times \begin{cases} -\sqrt{p_j}|j,L\rangle+\sqrt{q_j}|j,R\rangle & \text{: $j\geq 1$,}  \\ 
-\sqrt{p_0}|0,O\rangle+\sqrt{r_0}|0,R\rangle & \text{: $j=0$.} \end{cases}
\end{align}
Here 
\[C_R'= \sum_{j\geq 1} \frac{r_0q_1\cdots q_{j-1}}{p_1\cdots p_{j}}.  \]
\end{corollary}
In the case of a RW, the existence of finite number of self loops have little influence to the behavior of the RW itself. 
However as is seen in the next corollary, the existence of only one self loop changes localization property of the corresponding QW. 
\begin{corollary}
In a recurrent RW with $\mathcal{S}=\{0\}$, localization of the corresponding QW does not occur for any initial states in 
$\bs{\Psi}_0$ satisfying the assumptions (1) and (2). 
On the other hand, once we add an additional self loop at position $n>0$, 
then an appropriate choice of initial state from $\mathrm{span}\{\bs{\mathfrak{q}}_\pm (x_0): x_0\in \mathcal{B}\}^{\bot}$ 
so that $|\langle \bs{a}^{\bot,(n)}_j, \bs{\Psi}_0 \rangle|>0$ 
gives localization whose time averaged limit measure is described by 
\begin{equation}\label{seigo}
\overline{\mu}_\infty^{(\Psi_0)}(i) = \left|\sum_{j\geq 0}\langle \bs{a}^{\bot,(n)}_j, \bs{\Psi}_0 \rangle\right|^2\bs{\pi'}^{(n)}(i), 
\end{equation}
where 
\begin{align} 
\bs{\pi'}^{(n)}(j) &= 
	\frac{I_{\{ 0\leq j \leq n \}}}{1+{C_R^{(n)}}'} \times 
        \begin{cases}
	1 & \text{: $j=0$,} \\
        \frac{r_0q_1\cdots q_{j-1}}{p_1\cdots p_j}\times I_{\{n>1\}} & \text{: $1\leq j\leq n-1$,} \\
        \frac{r_0q_1\cdots q_{n-1}}{p_1\cdots p_{n-1}r_n}\times (1-p_n) & \text{: $j=n$,} 
	\end{cases}  \\       
\bs{a}^{\bot,{(n)}}_j &= (-1)^j\sqrt{\bs{\pi'}^{(n)}(j)}\times 
	\begin{cases}
	-\sqrt{p_0}|0,O\rangle+\sqrt{r_0}|0,R\rangle & \text{: $j=0$,} \\
	(-\sqrt{p_j}|j,L\rangle+\sqrt{q_j}|j,R\rangle) \times I_{\{n>1\}} & \text{: $1\leq j\leq n-1$,}  \\ 
	-\sqrt{p_n}|n,L\rangle+\sqrt{p_nq_n/r_n}|n,O\rangle & \text{: $j=n$.} 
	\end{cases}
\end{align}
Here 
\[{C_R^{(n)}}'= I_{\{n>1\}}\times \left\{
	\sum_{j=1}^{n-1} \frac{r_0q_1\cdots q_{j-1}}{p_1\cdots p_{j}}
        +\frac{r_0q_1\cdots q_{n-1}}{p_1\cdots p_{n-1}r_n}\times(1-p_n)\right\}+I_{\{n=1\}}\frac{r_0(1-p_1)}{r_1}.  
\]
\end{corollary}
We have found that the time averaged limit measure is described by the summation over 
all the orthogonal projections of the initial state onto each $\ell^2(A)$ eigenvector of $U$. 
The $\ell^2(V)$ eigenvectors of the corresponding RW give all the $\ell^2(A)$ eigenvectors in $\mathcal{H}^{(R)}$. 
In general, it is difficult to obtain all the $\ell^{2}(V)$ eigenvectors except 
the maximal eigenvector corresponding to the stationary measure of the RW. 
By using the available eigenvectors (including the stationary measure), 
we have given a lower bound of the time averaged limit measure and evaluate localization of the QW. 
In some cases, all the $\ell^2(V)$ eigenvectors of the RW corresponding to all the mass points of the spectral measure can be computed. 
For example, from a quantum probabilistic approach, in the case of $p_j=p$, $q_j=q$ and $r_j=r$ with $p_j+q_j+r_j=1$, 
the spectral measure of the Jacobi matrix is expressed by the free Meixner law \cite{IO}. 
In Ref. \cite{KOS}, by using this, the time averaged limit measure in this case is explicitly obtained. 
Moreover they give an expression for a complete orthonormal system of $\mathcal{H}^{(S)}$ in a structure of the Sierpinski gasket, 
and then a condition of the initial state for localization with a bounded support can be shown. 
\section{Discussion}
We decomposed the eigensystem of the QW into $\mathcal{H}^{(R)}$ induced by the RW and its orthogonal complement 
$\mathcal{H}^{(S)}$ whose eigenvalue is $-1$. 
We showed that the fundamental derivation of localization is the orthogonal projection of the initial state 
onto $\ell^2(A)$ eigenvectors which derives from $\ell^2(V)$ eigenvectors of the RW, and also 
onto the subspace $\mathcal{H}^{(S)}$. 
It was obtained that sufficient conditions for localization are expressed by recurrence properties of the RW. 
These are quenched results. 

Due to the infiniteness of the system, the distributivity of the time averaged limit measure is not ensured. 
Let $c_0$ be the summation of the time averaged limit measure over all the positions. Then how can we characterize the value $1-c_0$? 
From now on, we call the value $1-c_0$ ``missing value". 
For example, when $p_j=p$ $(j\geq 1)$, $=1$ $(j=0)$, and $q_j=q$ $(j\geq 1)$ with $p+q=1$, then according to \cite{CKS}, we see that 
the missing value appears at the following weak limit theorem. 
\begin{equation} \label{gosaku}
\frac{X_t}{t}\Rightarrow c_0\delta_0(x)+(1-c_0)\frac{x^2}{1-|p-q|} f_K(x:2\sqrt{pq})\;\;(t\to\infty), 
\end{equation}
where $f_K(x:s)$ is the Konno density function \cite{Konno1,Konno2} with parameter $s\in(0,1)$, $c_0=I_{\{p<q\}}(1-p/q)$ and 
``$\Rightarrow$" means the weak convergence. This is so called ``clean" quantum walk case. 
Now how about disordered quantum walk case? 
In the case of one defect on the doubly infinite line \cite{KLS}, 
the missing value, which spreads away ballistically, appears in the weak limit theorem which is expressed by a similar formulas of Eq. (\ref{gosaku}): 
the density is 
a convex combination of a delta measure at the origin 
corresponding to localization, and a product of a rational polynomial and the Konno density function. 
This result of \cite{KLS} suggests that a small number of perturbations generates edge state of the QW, and removes 
the missing value corresponding to bulk state ballistically to infinite places keeping a shape characterized by the Konno density functions 
as is the second term of RHS of Eq. (\ref{gosaku}). 
On the other hand, interestingly, Obuse and Kawakami \cite{OK} and Ahlbrecht {\it et al.} \cite{Hanover} suggest that 
a large number of perturbations, that is, 
an random environment, gives a characterization of the missing value by an exponential function named dynamical localization 
with not ballistic but sub-ballistic spreading. 

Finally, the quantum coins treated in this paper belong to a special class of unitary operator. 
The stochastic process behind this QW with this choice of quantum coins makes the computations be easy and gives a nice relationship 
of behaviors between the RW and the QW. 
What kind of walk exists behind the QW or the CMV matrix \cite{CGMV} with another choice of quantum coins such as \cite{Joye}? 
Applying the frame work in this paper to another kind of class of quantum coins is one of the interesting future problems. 

\end{document}